\documentclass[a4paper,conference]{IEEEtran}
\usepackage{times,amsmath,amssymb}
\usepackage{slashbox}
\usepackage{graphicx}
\usepackage{xcolor}
\usepackage{algorithm2e}
\usepackage{algorithmicx}
\usepackage{setspace}
\usepackage[export]{adjustbox}
\usepackage{multirow}
\usepackage{bm}
\usepackage[noend]{algpseudocode}
\usepackage{stmaryrd}
\usepackage{soul}

\begin{document}

\title{Bilinear Input Normalization for Neural Networks in Financial Forecasting}
\author{\IEEEauthorblockN{Dat Thanh Tran\IEEEauthorrefmark{1}, Juho Kanniainen\IEEEauthorrefmark{1}, Moncef Gabbouj\IEEEauthorrefmark{1}, Alexandros Iosifidis\IEEEauthorrefmark{2}}
\IEEEauthorblockA{\IEEEauthorrefmark{1}Department of Computing Sciences, Tampere University, Finland\\
\IEEEauthorrefmark{2}Department of Electrical and Computer Engineering, Aarhus University, Denmark\\
Email:\{thanh.tran, juho.kanniainen, moncef.gabbouj\}@tuni.fi, ai@ece.au.dk}\\

}

\maketitle

\begin{abstract}
Data normalization is one of the most important preprocessing steps when building a machine learning model, especially when the model of interest is a deep neural network. This is because deep neural network optimized with stochastic gradient descent is sensitive to the input variable range and prone to numerical issues. Different than other types of signals, financial time-series often exhibit unique characteristics such as high volatility, non-stationarity and multi-modality that make them challenging to work with, often requiring expert domain knowledge for devising a suitable processing pipeline. In this paper, we propose a novel data-driven normalization method for deep neural networks that handle high-frequency financial time-series. The proposed normalization scheme, which takes into account the bimodal characteristic of financial multivariate time-series, requires no expert knowledge to preprocess a financial time-series since this step is formulated as part of the end-to-end optimization process. Our experiments, conducted with state-of-the-arts neural networks and high-frequency data from two large-scale limit order books coming from the Nordic and US markets, show significant improvements over other normalization techniques in forecasting future stock price dynamics.    
\end{abstract}

\section{Introduction}\label{S:Intro}
Nowadays, the world economical and social developments and well-beings are heavily influenced by financial markets. People participate in financial activities, which promote the circulation of assets and developments of the world economy, with the ultimate goal of gaining economic benefits. Under this light, the success of the participants depends largely on the quality and quantity of information that they possess, as well as their ability to interpret these information for decision-making. Because of this, computational intelligence in finance, which utilizes modern computing methodologies to analyze financial markets for decision-making, has attracted many researchers and practitioners from both academia and industry. Representative topics under this discipline include stock market forecasting \cite{tran2018temporal, zhang2019deeplob}, algorithmic trading \cite{nuti2011algorithmic, hu2015application}, risk assessment \cite{khandani2010consumer, galindo2000credit}, asset pricing \cite{cochrane1996cross, lettau2020estimating}, and portfolio allocation and optimization \cite{demiguel2009generalized, ban2018machine}. Among these objectives, a substantial amount of research efforts has been dedicated to prediction and forecasting since financial decision-making, for the most part, depends on reliable projections about the future. 

There are two common approaches, namely fundamental analysis \cite{thomsett2006getting} and technical analysis \cite{murphy1999technical}, which are currently adopted in predicting future market behaviors. In fundamental analysis, valuation techniques take into account different economic indicators that reflect and affect the market movements to establish long-term views on the development of a financial entity. On the other hand, in technical analysis, it is generally believed that the prices themselves already encompass all factors that affect the market dynamics. For this reason, technical analysts construct forecasting models based on series of historical transactions with the assumption that history tends to repeat itself \cite{murphy1999technical}, and the underlying processes, which generate the observed series, can be captured by mathematical or computational models.   

Although financial time-series forecasting has been extensively studied over the past decades with a large body of literature dedicated to tackling specific problems, there are still many challenges in processing and analyzing data derived from financial markets, especially those coming from high-frequency intra-day activities. Over time, the development of internet technologies, database systems and electronic trading platforms have enabled us to collect a vast amount of digital footprints of the financial market. Enormous volumes of data, while ensuring statistical significance of any analysis, also create a great computational challenge when building financial prediction models. The computational aspect is especially critical for trading applications that take advantage statistical arbitrage, which usually exists in very short time before market correction \cite{avellaneda2010statistical}. Another challenge posed by financial time-series comes from the fact that they are usually complex, noisy, nonlinear and nonstationary in nature, which leads to difficulties not only in modeling but also in preprocessing.  

Techniques for financial time-series prediction fall into two categories: traditional statistical models and machine learning models. In the stochastic model based approach, linear relationship is often assumed between the independent variables. Representative tools in this category include autoregressive integrated moving average (ARIMA) and its variants or generalized autoregressive conditional heteroskedasticity (GARCH) \cite{engle1982autoregressive}, to name a few. While stochastic models often possess nice theoretical properties, the underlying assumption is often too strong, leading to poor generalization performance in real-world data. On the other hand, machine learning models, which make no prior statistical or structural assumption, are often capable of modeling complex nonlinear relationships among the independent factors and the prediction targets. For this reason, machine learning models often generalize better than stochastic models in many forecasting scenarios \cite{kane2014comparison, qian2017financial}.  

Among different types of machine learning models, neural networks are the leading solutions for many financial forecasting problems nowadays \cite{korczak2017deep, tran2018temporal, zhang2019deeplob, tsantekidis2017forecasting, dingli2017financial}. The majority of these solutions were adopted from computer vision (CV) and natural language processing (NLP) applications where neural networks have demonstrated unprecedented successes in the last decade. Despite the fact that future market prediction based on historical time-series can be casted as a pattern recognition problem similar to those encountered in CV and NLP, thus can be treated in some degree of success with tools from CV and NLP, the unique characteristics of financial data make the market prediction tasks fundamentally different and require special treatments. The majority of problems targeted in CV and NLP concern solving cognitive tasks in which the data is intuitive and well-understood by normal human beings, such as recognition of objects or understanding natural languages. On the other hand, historical financial phenomena even require human experts to recognize or interpret, not to mention speculating about the future. In addition, images, videos or speeches, for example, are well-behaved signals in the sense that the value range and variances are known and can be easily processed without losing the essential information within them, while financial time-series are highly volatile and often exhibit concept drift phenomena \cite{clements2004forecasting, hatemi2008tests}, i.e., dynamic changes in the relationship between independent and target variables over time. Because of this, data preprocessing is an important procedure when working with financial time-series.            

Among many preprocessing steps, data normalization, which is one of the most essential steps before building a machine learning model, aims at transforming input variables into a common range to avoid the potential bias induced by large numbers. For deep neural networks, improperly normalized data can easily lead to numerical issues with the gradient updates. In literature, there  are many normalization methods such as z-score normalization, min-max normalization, pareto scaling, power transformation, to name a few \cite{singh2020investigating}. These normalization methods utilize global data statistics, such as the mean, standard deviation or maximum value to transform the data. For financial time-series, especially those covering long periods, replacing global statistics with local statistics computed over the recent history is a common practice to avoid the problem of potential regime shifts in which recent observations have significantly different value range than past observations. To deal with this phenomenon, several sophisticated methods have been proposed, for example \cite{shao2015self, nayak2014impact}.  

While many static normalization schemes have been developed as described above, we are only aware of one prior work \cite{passalis2019deep} that proposed an adaptive method for input time-series. Different from static approaches, an adaptive data-driven method transforms raw input data using statistics that are identified and learned via optimization. That is, the step is implemented as the first layer in a computation graph, with all parameters jointly estimated using stochastic gradient descent. In fact, one of the reasons that make neural nets work so well is the fact that they are estimated in an end-to-end manner, being able to learn data-dependent transformations. Thus, we argue that the normalization step for input time-series should also be learned in the same end-to-end manner when employing neural networks in financial forecasting.    

In this paper, we propose Bilinear Input Normalization (BiN), a neural network layer that takes into account the bimodal nature of multivariate time-series, and performs input data transformation using parameters that are jointly estimated with other parameters in the network. The preliminary results of this work was presented in \cite{tran2020data}, which includes limited analysis and empirical evaluation of BiN for Temporal Attention Augmented Bilinear Layer (TABL) networks. In this paper, we provide more detailed, in-depth presentation and discussion of the proposed method, as well as extensive experiments demonstrated with another state-of-the-arts (SoTA) architecture in financial forecasting using stock market data from two different markets (US and Nordic).

The remainder of the paper is organized as follows. In Section \ref{related-works}, we review related works in data normalization methods, with a focus on normalization schemes for neural networks. Section \ref{method} describes in details the motivation and operations of the Bilinear Input Normalization layer. In Section 4, we provide basic information regarding limit order books and describe the problem of predicting stock mid-price dynamics using limit order book data, which is followed by the experimental setup, dataset description, the results and our analysis. Section \ref{conclusions} concludes our work. 
 
\section{Related Work}\label{related-works}

Normalization is a scaling or transformation operation, usually in a linear manner, to ensure a uniform value range between different data dimensions, reducing the effects of dominant values and outliers \cite{garcia2015data}. Perhaps, the most common normalization method is z-score normalization, which centers the data around the origin with unit standard deviation. There are also works that only center the data, without the scaling step as in z-score normalization. The steps in Pareto scaling \cite{noda2008scaling} are similar to z-score normalization, except for the division of standard deviation instead of the variance. A generalization of z-score normalization is the variance stability scaling method \cite{van2006centering}, which multiplies the z-score standardized data with the ratio between the mean and standard deviation of the data. Power transformation is another normalization method employing the mean statistic to reduce the effects of heteroscedasticity \cite{kvalheim1994preprocessing}. Besides data's mean and variance, minimum, maximum and median values are also utilized in normalization, such as min-max normalization, median and median absolute deviation normalization. For interested readers, we refer to the analysis of different static data normalization techniques in machine learning models \cite{singh2020investigating}.    

The term data normalization is often understood as the operation that preprocesses raw data, i.e., input data. However, in neural networks, normalization operation is also popular in hidden layers. This is due to the fact that different layers in a deep network can encounter significant input distribution shift during stochastic gradient updates. Normalization operation can be used to help stablize and improve the training process. Batch Normalization (BN) was proposed for Convolutional Neural Networks such a purpose \cite{ioffe2015batch}. Since stochastic gradient descent only operates in a mini-batch manner, the mini-batch mean and variance are accumulated in a moving average style to estimate the global mean and variance in BN. After subtracting the mean and dividing by the variance, BN also learns to scale and shift the hidden representations. Instead of the mini-batch statistics, Instance Normalization \cite{ulyanov2016instance} uses sample-level statistics, and learns how to normalize each image so that its contrast matches with that of a predefined style image in the visual style transfer problems. Both BN and IN were originally proposed for visual data, although BN has also been widely used in NLP.

Both BN and IN are adaptive data-driven normalization schemes. However, they were proposed to normalize the hidden representations, and they are not commonly used for input normalization. Regarding adaptive input normalization method for time-series, we are only aware of the work in \cite{passalis2019deep}, which formulated a 3-stage normalization procedure called Deep Adaptive Input Normalization (DAIN). Since DAIN is directly related to our proposed method, we describe DAIN in more details here.  

In this paper, let us denote the collection of $N$ multivariate series as $\{\mathbf{X}^{(n)} \in \mathbb{R}^{D \times H}\;|n=1, \dots, N\}$, where $D$ denotes the number of univariate series and $H$ denotes the temporal length of each series. Here $D$ and $H$ are also referred to as the feature and temporal dimensions, respectively.  In addition, we denote the $h$-th column of $\mathbf{X}^{(n)}$ as $\mathbf{c}_h^{(n)} \in \mathbb{R}^{D}$, which is the representation of the series at the time index $h$. We also refer to $\mathbf{c}_h^{(n)}$ as the $h$-th temporal slice. The first step of DAIN is to shift every temporal slice in $\mathbf{X}^{(n)}$ as follows:

\begin{equation}\label{eq1}
\begin{aligned}
	&\bar{\mathbf{c}}^{(n)} = \frac{1}{H} \sum_{h=1}^{H} \mathbf{c}_h^{(n)} \\
	& \mathbf{y}_h^{(n)} =  \mathbf{c}_h^{(n)} - \mathbf{W}_{a} \bar{\mathbf{c}}^{(n)}, \quad \forall h=1, \dots, H
\end{aligned}
\end{equation}
where $\mathbf{W}_{a} \in \mathbb{R}^{D \times D}$ is a learnable weight matrix that estimates the amount of shifting from the mean temporal slice ($\bar{\mathbf{c}}^{(n)}$) calculated from each series. 

\begin{figure*}[t]
\centering
        \includegraphics[width=1.0\textwidth]{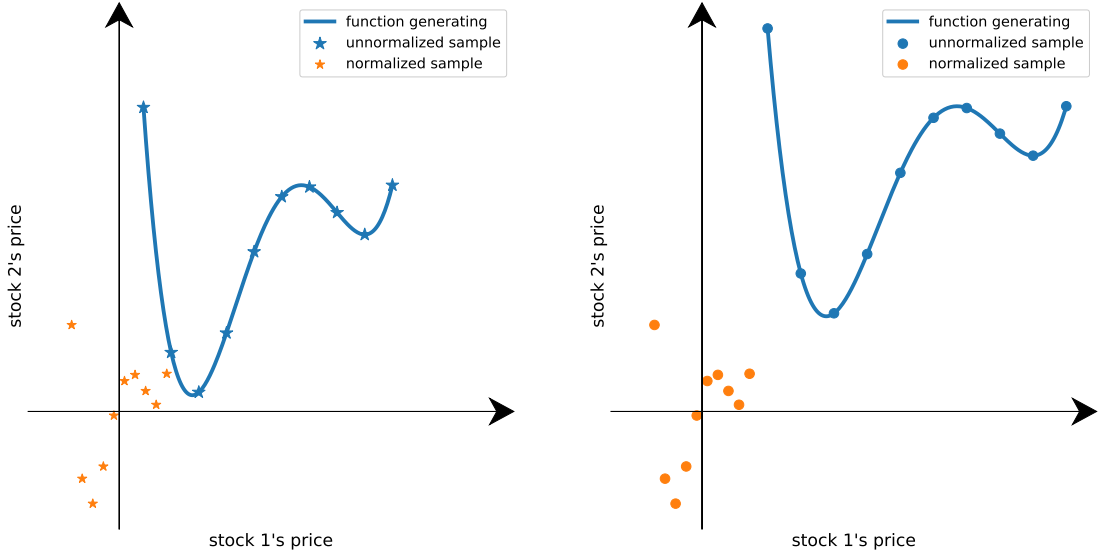}%
	\caption{Illustration of the effect of normalization along temporal mode. Here we consider two samples $\mathbf{X}^{(n_1)}$ and $\mathbf{X}^{(n_2)}$ on the left and right sides, respectively, each of which contains the opening prices of two stocks for 10 consecutive days, thus the multivariate series has dimensions $2\times 10$. The continuous line represents the function governing the relationship between two stocks and the scatter plots represent the prices that we observe (our samples). We can see that compared to prices at $\mathbf{X}^{(n_1)}$, the price range at the time of $\mathbf{X}^{(n_2)}$ has shifted for both stocks but their relationship is similar (the relative arrangement of points in $2$-dimensional space is similar, but with different amounts of spread). After the normalization step (here we simply demonstrate with scaling factor of one and no shifting), the arrangements of normalized points are positioned at the same place in this $2$-dimensional space, with similar spreads.}
        \label{f1}
\end{figure*}

After shifting, the intermediate representation $\mathbf{y}_h^{(n)}$ is then scaled as follows:

\begin{equation}\label{eq2}
\begin{aligned}
	& \bm{\sigma}^{(n)} = \sqrt{\frac{1}{H} \sum_{h=1}^{H} \big(\mathbf{y}_h^{(n)} \odot \mathbf{y}_h^{(n)}\big)}\\
	& \mathbf{z}_h^{(n)} =  \mathbf{y}_h^{(n)} \varoslash \big(\mathbf{W}_b \bm{\sigma}^{(n)}\big), \quad \forall h=1, \dots, H
\end{aligned}
\end{equation}
where $\mathbf{W}_{b} \in \mathbb{R}^{D \times D}$ is another weight matrix that estimates the amount of scaling from the standard deviation ($\bm{\sigma}^{(n)}$), which is computed from $H$ temporal slices. In Eq. (\ref{eq2}), the square-root operator is applied element-wise; $\odot$ and $\varoslash$ denote the element-wise multiplication and division, respectively. 

The final step in DAIN is gating, which is used as a type of attention mechanism to suppress irrelevant features:

\begin{equation}\label{eq3}
\begin{aligned}
\bar{\mathbf{z}}^{(n)} &= \frac{1}{H} \sum_{h=1}^{H} \mathbf{z}_h^{(n)}\\
\bm{\gamma}^{(n)} & = \mathrm{sigmoid}\big(\mathbf{W}_{c} \bar{\mathbf{z}}^{(n)} + \mathbf{W}_d \big) \\
\mathbf{t}_h^{(n)} &=  \mathbf{z}_h^{(n)} \odot \bm{\gamma}^{(n)} , \quad \forall h=1, \dots, H
\end{aligned}
\end{equation}
where $\mathbf{W}_c \in \mathbb{R}^{D\times D}$ and $\mathbf{W}_d \in \mathbb{R}^{D}$ are two weight matrices to learn the gating function. 

The output of DAIN is, thus, $\mathbf{T}^{(n)} = [\mathbf{t}_1^{(n)}, \dots, \mathbf{t}_H^{(n)}] \in \mathbb{R}^{D \times H}$, which is the normalized series having the same size as the input series $\mathbf{X}^{(n)}$. Since the normalization scheme of DAIN contains several processing steps with nonlinear operations, stochastic updates in DAIN are sensitive to the learning rate. For this reason, the authors in \cite{passalis2019deep} used three different learning rates for the parameters associated with three computational steps in DAIN. As we will see in the next section, our normalization scheme is more intuitive for time-series while requiring fewer computation and parameters. In addition, since our normalization scheme only relies on linear operations, it is robust with respect to the learning rates that are normally adopted to train the network under consideration.   

\section{Adaptive Input Normalization with Bilinear Normalization Layer}\label{method}

The proposed BiN layer formulation shares some similarities with DAIN and IN in the sense that we also propose to take advantage of sample-level statistics when learning to transform the input series. More specifically, the basic statistics, which are used to normalize each input sample, were calculated independently for each sample. There are also global parameters that are shared between samples in BiN. In this way, our formulation (as well as DAIN and IN) is different from BN, which utilizes global statistics estimated from the whole dataset to normalize every sample. For BN and IN, both methods were not proposed to work as an input normalization scheme for time-series, but to work with higher-order tensors in hidden layers of convolutional neural networks, which have different semantic structure than multivariate time-series. We are also not aware of any work that utilizes BN and IN for input data normalization, especially for time-series. The main difference between the proposed method and DAIN is that BiN is formulated to jointly learn to transform the input samples along both temporal and feature dimensions, taking into account the bimodal nature of multivariate time-series, while DAIN only works along the temporal dimension.

In order to better understand our motivation in taking into consideration the bimodal nature of multivariate time-series, let us take an example in predicting the opening value of NASDAQ-100 index of a day based on the historical opening prices of 100 constituent companies in the last 10 days. In this case, each input sample $\mathbf{X}^{(n)}$ has dimensions of $100\times 10$. On one hand, we can consider that each $\mathbf{X}^{(n)}$ is represented by a set of $10$ features ($10$ columns of $\mathbf{X}^{(n)}$), each of which has $100$ dimensions, representing the snapshot of the opening prices of $100$ constituent companies in NASDAQ-100. Thus, the mean value and variance of this set, also of $\mathbf{X}^{(n)}$, would represent the average opening prices and their volatility of $100$ companies in the last $10$ days. On the other hand, we can also consider that each $\mathbf{X}^{(n)}$ is represented by a set of $100$ univariate series, each of which contains opening prices of a company over $10$ consecutive days. Therefore, the mean value and variance of this set, also of $\mathbf{X}^{(n)}$, would represent the mean and variance of the NASDAQ-100 equal weighted index\footnote{This means that each constituent company contributes 1\%, without taking into account market capitalization. For example QQQE is an ETF that tracks NASDAQ-100 with equal weights} during the last $10$ days. In our example, both ways of considering $\mathbf{X}^{(n)}$ and the corresponding statistics are valid and meaningful. Each gives a different interpretation of the data contained in $\mathbf{X}^{(n)}$, as well as the underlying assumption about elements being normally distributed in the set representing $\mathbf{X}^{(n)}$. Because of this, the proposed normalization layer utilizes and combines statistics from both views in order to transform the multivariate series.  

The proposed layer normalizes along the temporal dimension as follows:
\begin{subequations}
	\label{eq5}
	\begin{align}
		 \bar{\mathbf{c}}^{(n)}  &= \frac{1}{H} \sum_{h=1}^{H} \mathbf{c}_h^{(n)}\label{eq5.1}\\
		 \bm{\sigma}_2^{(n)} &= \sqrt{\frac{1}{H} \sum_{h=1}^{H}\big(\mathbf{c}_h^{(n)} - \bar{\mathbf{c}}^{(n)}\big) \odot \big(\mathbf{c}_h^{(n)} - \bar{\mathbf{c}}^{(n)}\big)} \label{eq5.2}\\
		 \mathbf{a}_h^{(n)} &= \bm{\gamma}_2 \odot \big((\mathbf{c}_h^{(n)} - \bar{\mathbf{c}}^{(n)}) \varoslash \bm{\sigma}_2^{(n)}\big) + \bm{\beta}_2, \quad \forall h=1, \dots, H \label{eq5.3}\\
		\mathbf{A}^{(n)} &= [\mathbf{a}_1^{(n)}, \dots, \mathbf{a}_h^{(n)}, \dots, \mathbf{a}_H^{(n)}] \in \mathbb{R}^{D\times H}
	\end{align}
\end{subequations}
where $\bm{\gamma}_2 \in \mathbb{R}^{D}$ and $\bm{\beta}_2 \in \mathbb{R}^D$ are two parameters of BiN that are optimized during stochastic gradient descent.   

After the computation steps in Eq. (\ref{eq5}), we obtain an intermediate series $\mathbf{A}^{(n)}$ that has been normalized in the temporal dimension. Basically, in Eq. (\ref{eq5}), given an input series $\mathbf{X}^{(n)}$, BiN first computes the mean temporal slice (column) $\bar{\mathbf{c}}^{(n)} \in \mathbb{R}^{D}$ and its standard deviation $\bm{\sigma}_2^{(n)} \in \mathbb{R}^{D}$ as in Eq. (\ref{eq5.1}, \ref{eq5.2}), which are then used to standardize each temporal slice of the input before applying element-wise scaling (using $\bm{\gamma}_2$) and shifting (using $\bm{\beta}_2$) as in Eq. (\ref{eq5.3}). While the standardizing step is independent for each sample in the training set, last shifting and scaling parameters are shared between all samples. Here we use the subscript ($2$) in $\bm{\sigma}_2^{(n)}$, $\bm{\gamma}_2$ and $\bm{\beta}_2$ to indicate that they are associated with the second dimension, i.e., the temporal dimension, of the multivariate series.

In order to interpret the effects of Eq. (\ref{eq5.1}), (\ref{eq5.2}), and (\ref{eq5.2}), we can take the same approach as the example given for NASDAQ-100 previously. That is, the input series $\mathbf{X}^{(n)}$ can be viewed as the set $\mathcal{T}^{(n)}$ consisting of $H$ temporal slices, i.e., a set consisting of $H$ points in a $D$-dimensional space. The first part in Eq. (\ref{eq5.3}), i.e. $(\mathbf{c}_h^{(n)} - \bar{\mathbf{c}}^{(n)}) \varoslash \bm{\sigma}_2^{(n)}$, moves this set of points around the origin and as well as controlling their spread while keeping their arrangement pattern similarly. If we have two input series $\mathbf{X}^{(n_1)}$ and $\mathbf{X}^{(n_2)}$ with the corresponding sets $\mathcal{T}^{(n_1)}$ and $\mathcal{T}^{(n_2)}$ spreading and lying in two completely different areas of this $D$-dimensional space but have the same arrangement pattern, without the alignment performed by the first part of Eq. (\ref{eq5.3}), we cannot effectively capture the linear or nonlinear\footnote{Nonlinear patterns can be estimated by several piece-wise linear patterns (using more than one linear projections such as more than one convolution filters)} arragement patterns that are similar between the two series when using, for example, a 1D convolution filter that strides along the temporal dimension as often encountered in CNN architectures for time-series. We illustrate our example in Figure \ref{f1}. Here we should note that although BiN applies additional scaling and shifting in Eq. (\ref{eq5.3}) after the alignment, the values of $\bm{\gamma}_2$ and $\bm{\beta}_2$ are the same for every input series, thus the points of the set $\mathcal{T}^{(n_1)}$ and $\mathcal{T}^{(n_2)}$ are still centered at the same point and having approximately similar spreads. Since $\bm{\gamma}_2$ and $\bm{\beta}_2$ are optimized together with other network's parameters, they enable BiN to manipulate the aligned distributions of $\mathcal{T}^{(n)}$ to match with the statistics of other layers. 

While the effect of non-stationarity in the temporal mode are often visible and has been heavily studied, its effects when considered from the feature dimension perspective are less obvious. To see this, let us now view the series $\mathbf{X}^{(n)}$ as the set $\mathcal{F}^{(n)}$ of $D$ points (its $D$ rows) in a $H$-dimensional space. Let us also take the previous scenario where two series, $\mathbf{X}^{(n_1)}$ and $\mathbf{X}^{(n_2)}$, have $\mathcal{T}^{(n_1)}$ and $\mathcal{T}^{(n_2)}$ scattered in different regions of a $D$-dimensional co-ordinate system (viewed under the temporal perspective) before the normalization step in Eq. (\ref{eq5}). When $\mathcal{T}^{(n_1)}$ and $\mathcal{T}^{(n_2)}$ are very far away viewed from the feature perspective, these two series are also likely to possess $\mathcal{D}^{(n_1)}$ and $\mathcal{D}^{(n_2)}$ which are distributed in two different regions of a $H$-dimensional space, although having very similar arrangement. This scenario also prevents a convolution filter that strides along the feature dimension to effectively capture the prominent linear/nonlinear patterns existing in the feature dimension of all input series. For this reason, our proposed normalization scheme also normalizes the input series along the feature dimension as follows:

\begin{subequations}
	\label{eq6}
	\begin{align}
		 \bar{\mathbf{r}}^{(n)}  &= \frac{1}{D} \sum_{d=1}^{D} \mathbf{r}_d^{(n)}\label{eq6.1}\\
		 \bm{\sigma}_1^{(n)} &= \sqrt{\frac{1}{D} \sum_{d=1}^{D}\big(\mathbf{r}_d^{(n)} - \bar{\mathbf{r}}^{(n)}\big) \odot \big(\mathbf{r}_d^{(n)} - \bar{\mathbf{r}}^{(n)}\big)} \label{eq6.2}\\
		 \mathbf{b}_d^{(n)} &= \bm{\gamma}_1 \odot \big((\mathbf{r}_d^{(n)} - \bar{\mathbf{r}}^{(n)}) \varoslash \bm{\sigma}_1^{(n)}\big) + \bm{\beta}_1, \quad \forall d=1, \dots, D \label{eq6.3}\\ 
		 \mathbf{B}^{(n)} &= \begin{bmatrix} \mathbf{b}_1^{(n)} \\ \vdots \\ \mathbf{b}_d^{(n)} \\ \vdots \\ \mathbf{b}_D^{(n)}\end{bmatrix} \in \mathbb{R}^{D\times H} \label{eq6.4}
	\end{align}
\end{subequations}
where $\mathbf{r}_d^{(n)} \in \mathbb{R}^{H}$ denotes the $d$-th row of $\mathbf{X}^{(n)}$. In addition, $\bm{\gamma}_1 \in \mathbb{R}^{H}$ and $\bm{\beta}_1 \in \mathbb{R}^{H}$ are two learnable weights.

After computing the steps in Eq. (\ref{eq6}), we obtain another intermediate series $\mathbf{B}^{(n)}$ that has been normalized in the feature dimension. 

Finally, BiN linearly combines the intermediate normalized series obtained from Eq. (\ref{eq5}) and (\ref{eq6}) to generate the output $\mathbf{T}^{(n)} \in \mathbb{R}^{D\times H}$:
\begin{equation}\label{eq7}
	\mathbf{T}^{(n)} = \lambda_a \mathbf{A}^{(n)} + \lambda_b \mathbf{B}^{(n)} 
\end{equation} 
where $\lambda_a \in \mathbb{R}$ and $\lambda_b \in \mathbb{R}$ are two learnable scalars, which enable BiN to weigh the importance of temporal and feature normalization. Here we should note that $\lambda_a$ and $\lambda_b$ are constrained to be non-negative. This constraint is achieved during stochastic optimization by setting the value (of $\lambda_a$ or $\lambda_b$) to $0$ whenever the updated value is negative.

\section{Experiments}\label{experiments}

\subsection{Limit Order Book}
In finance, a limit order is a type of trade order to buy or sell a fixed number of shares with a specified price. In a buy (bid) limit order, the trader specifies the number of shares and the maximum price per share of the stock that he or she is willing to pay. On the contrary, for a sell (ask) limit order, the trader must specifies the number of shares and the minimum share price that he or she wants to sell. The two types of limit order form the two sides of the limit order book (LOB): the bid and the ask sides. The limit orders are sorted such that the ones with the highest bid price are on top of the bid side and the ones with the lowest ask price are on top of the ask side. Whenever the best ask price is equal or lower than the best bid price, those orders are executed and removed from the LOB.  

Since the LOB contains all the transactions related to a stock, it reflects the current supply and demand of the stock at different price levels. In literature, there are numerous researches that take advantage of the LOB data and formulate different research questions such as order flow distribution, price jumps, random walk nature of prices, stochastic models of limit orders, to name a few \cite{siikanen2017limit, siikanen2017drives, bouchaud2004fluctuations, cont2013price, makinen2019forecasting}. One of the problems related to the LOB that are heavily studied using machine learning methods is the problem of forecasting future mid-price movements. Mid-price, at any point in time, is the average value between the best-bid and best-ask prices. This quantity is a virtual price since no trade can happen at the current mid-price. Since the movements of mid-price reflect the changes in market dynamics, they are considered as important events to forecast. In order to benchmark performances of BiN, we conducted experiments using two different LOB datasets coming from two different markets: Nordic and US markets. 

\subsection{Experiments using Nordic data}

\subsubsection{Dataset and Experimental Setup}

FI-2010 \cite{ntakaris2018benchmark} is a large scale, publicly available Limit Order Book (LOB) dataset, which contains buy and sell limit order information (the prices and volumes) over $10$ business days from $5$ Finnish stocks traded in Helsinki Stock Exchange (operated by NASDAQ Nordic). At each order event (a point in time), the dataset contains the prices and volumes from the top $10$ best-bid and best-ask orders of both sides, leading to a $40$-dimensional vector representation. The authors of this dataset provided the labels (up, down, stationary) for the mid-price movements in the next $\{10, 20, 30, 50, 100\}$ order events. Since the majority of existing research results were reported for prediction horizons in the set $H = \{10, 20, 50\}$, we also conducted experiments with these values. Interested readers can read more about the FI-2010 dataset in \cite{ntakaris2018benchmark}.

For the FI-2010 dataset, we followed the same experimental setup proposed in \cite{tran2018temporal}, which is widely used to benchmark the performances of deep neural networks in this task. Under this setting, data of the first $7$ days was used to train the models, and the last $3$ days were used for evaluation purposes. In this first set of experiments, we evaluated BiN in combination with the Temporal Attention augmented Bilinear Layer (TABL) network, which is one of the SoTA neural networks in FI-2010 dataset \cite{tran2018temporal}. Since TABL architectures also take advantage of the bimodal nature of the time-series, BiN is expected to ideally complement TABL networks. To enable comparisons with prior works, the best performing architecture C(TABL) reported in \cite{tran2018temporal} was adopted in our experiments. For this architecture, the input time-series were constructed from $10$ most recent order events. As we mentioned above, since at each order event, the LOB is represented by a $40$-dimensional vector, each input series that is fed to C(TABL) has dimensions of $40\times 10$. All C(TABL) networks were trained with ADAM optimizer for $80$ epochs, with an initial learning rate of $0.001$, which was reduced by a factor of $10$ at epoch $11$ and $71$. Weight decay ($0.0001$) and max-norm constraint ($10.0$) were used for regularization.  

Accuracy, average Precision, Recall and F1 are reported as the performance metrics. Since FI-2010 is an imbalanced dataset, average F1 measure is considered as the main performance metric for FI-2010 following prior conventions \cite{tran2018temporal}. Here we should note that we used no validation set for FI-2010, and simply used the F1 score measured on the train set for validation purposes. Each experiment was run $5$ times and the median value measured on the test set is reported. 

\subsubsection{Experiment Results}

\begin{table}[t!]
	\begin{center}
		\caption{Experiment Results. Methods without any indication of normalization method means that z-score normalization was applied. Bold-face numbers denote the best F1 measure between the same model using different normalization methods. }\label{t1}
		\resizebox{\linewidth}{!}{
			\begin{tabular}{|c|c|c|c|c|}
				\multicolumn{5}{c}{} \\ \hline
				\textbf{Models}		& \textbf{Accuracy \%} 	& \textbf{Precision \%} & \textbf{Recall \%}	& \textbf{F1 \%} 		\\ \hline \hline
				\multicolumn{5}{|c|}{\textit{Prediction Horizon $H=10$}} \\ \hline
				CNN\cite{tsantekidis2017forecasting}		& -			& $50.98$	&$65.54$	& $55.21$	\\ \hline
				LSTM\cite{tsantekidis2017using}	& -			& $60.77$	&$75.92$	& $66.33$	\\ \hline \hline
				C(BL) \cite{tran2018temporal}	& $82.52$	& $73.89$	&$76.22$	& $75.01$ \\ \hline
				DeepLOB \cite{zhang2019deeplob}	& $84.47$	& $84.00$	&$84.47$	& $83.40$ \\ \hline \hline
				DAIN-MLP \cite{passalis2019deep}	& -	& $65.67$	&$71.58$	& $68.26$ \\ \hline
				DAIN-RNN \cite{passalis2019deep}	& -	& $61.80$	&$70.92$	& $65.13$ \\ \hline \hline
				C(TABL)	\cite{tran2018temporal} & $84.70$	& $76.95$	&$78.44$	& $77.63$ \\ \hline 
				BN-C(TABL)  & $79.20$	& $68.48$	&$72.36$	& $66.87$ \\ \hline
				BiN-C(TABL)	& $86.87$	& $80.29$	&$81.84$	& $\mathbf{81.04}$ \\ \hline \hline
				\multicolumn{5}{|c|}{\textit{Prediction Horizon $H=20$}} \\ \hline
				CNN\cite{tsantekidis2017forecasting}		& -			& $54.79$	&$67.38$	& $59.17$	\\ \hline
				LSTM\cite{tsantekidis2017using}	& -			& $59.60$	&$70.52$	& $62.37$	\\ \hline 
				C(BL) \cite{tran2018temporal}	& $72.05$	& $65.04$	&$65.23$	& $64.89$ \\ \hline 
				DeepLOB	\cite{zhang2019deeplob} & $74.85$	& $74.06$	&$74.85$	& $72.82$ \\ \hline \hline
				DAIN-MLP \cite{passalis2019deep}	& -	& $62.10$	&$70.48$	& $65.31$ \\ \hline
				DAIN-RNN \cite{passalis2019deep}	& -	& $59.16$	&$68.51$	& $62.03$ \\ \hline \hline
				C(TABL)	\cite{tran2018temporal} & $73.74$	& $67.18$	&$66.94$	& $66.93$ \\ \hline
				BN-C(TABL)  & $70.70$	& $63.10$	&$63.78$	& $63.43$ \\ \hline
				BiN-C(TABL)	& $77.28$	& $72.12$	&$70.44$	& $\mathbf{71.22}$ \\ \hline \hline 
				\multicolumn{5}{|c|}{\textit{Prediction Horizon $H=50$}} \\ \hline
				CNN\cite{tsantekidis2017forecasting}		& -			& $55.58$	&$67.12$	& $59.44$	\\ \hline
				LSTM\cite{tsantekidis2017using}	& -			& $60.03$	&$68.58$	& $61.43$	\\ \hline 
				C(BL) \cite{tran2018temporal} 	& $78.96$	& $77.85$	&$77.04$	& $77.40$ \\ \hline
				DeepLOB \cite{zhang2019deeplob}	& $80.51$	& $80.38$	&$80.51$	& $80.35$ \\ \hline \hline
				C(TABL) \cite{tran2018temporal}	& $79.87$	& $79.05$	&$77.04$	& $78.44$ \\ \hline
				BN-C(TABL)  	& $77.16$	& $75.70$	&$75.04$	& $75.34$ \\ \hline
				BiN-C(TABL)	& $88.54$	& $89.50$	&$86.99$	& $\mathbf{88.06}$ \\ \hline
			\end{tabular}
		}
	\end{center}
\end{table}

Table \ref{t1} shows the experiment results in three prediction horizons $H=\{10, 20, 50\}$ of C(TABL) networks using Batch Normalization and BiN, in comparison with existing results. Here we should note that the data provided in FI-2010 has been anonymized, i.e., the prices and volumes of orders were normalized. For those results reported in Table \ref{t1} without any indication of the normalization method, it means that z-score normalization was applied. In addition, we attempted to evaluate DAIN using the C(TABL) architecture on FI-2010 dataset, however, we could not achieve reasonable performances since this normalization strategy requires extensive tuning of three different learning rates for different computation steps. Besides, in the original paper \cite{passalis2019deep}, DAIN was only applied to MLP and RNN networks. For this reason, we report the original results of DAIN using MLP and RNN in Table \ref{t1}. In the experiments using US data, we did obtain reasonable results with DAIN and comparisons with DAIN are made in Section \ref{us-experiments}.  

\begin{table}[t!]
	\begin{center}
		\caption{Improvement comparisons between BiN-C(TABL) versus BiN-B(TABL)}\label{t3}
		\resizebox{\linewidth}{!}{
			\begin{tabular}{|c|c|c|c|c|}
\multicolumn{5}{c}{} \\ \hline
\textbf{Models}		& \textbf{Accuracy \%} 	& \textbf{Precision \%} & \textbf{Recall \%}	& \textbf{F1 \%} 		\\ \hline \hline
\multicolumn{5}{|c|}{\textit{Prediction Horizon $H=10$}} \\ \hline
B(TABL)	\cite{tran2018temporal}	& $78.91$	& $68.04$	&$71.21$	& $69.20$ \\ \hline
C(TABL)	\cite{tran2018temporal} & $84.70$	& $76.95$	&$78.44$	& $77.63$ \\ \hline \hline
BiN-B(TABL) & $86.92$ 	& $80.43$	&$81.82$	& $\mathbf{81.10}$ \\ \hline
BiN-C(TABL)	& $86.87$	& $80.29$	&$81.84$	& $81.04$ \\ \hline \hline
\multicolumn{5}{|c|}{\textit{Prediction Horizon $H=20$}} \\ \hline
B(TABL) \cite{tran2018temporal}       & $70.80$       & $63.14$       &$62.25$        & $62.22$ \\ \hline \hline
C(TABL)	\cite{tran2018temporal} & $73.74$	& $67.18$	&$66.94$	& $66.93$ \\ \hline \hline
BiN-B(TABL) & $77.54$	& $72.56$	&$70.22$	& $\mathbf{71.29}$ \\ \hline
BiN-C(TABL)	& $77.28$	& $72.12$	&$70.44$	& $71.22$ \\ \hline \hline 
\multicolumn{5}{|c|}{\textit{Prediction Horizon $H=50$}} \\ \hline
B(TABL) \cite{tran2018temporal} & $75.58$       & $74.58$       &$73.09$        & $73.64$ \\ \hline
C(TABL) \cite{tran2018temporal}	& $79.87$	& $79.05$	&$77.04$	& $78.44$ \\ \hline \hline
BiN-B(TABL) & $88.44$	& $89.36$	&$86.92$	& $87.96$ \\ \hline
BiN-C(TABL)	& $88.54$	& $89.50$	&$86.99$	& $\mathbf{88.06}$ \\ \hline
\end{tabular}
}
	\end{center}
\end{table}

It is clear that our proposed BiN layer (BiN-C(TABL)) when used to normalize the input data yielded significant improvements over BN and z-score normalization when applied to the same network. The improvements are obvious for all prediction horizons. Especially, for the longest horizon $H=50$, BiN enhanced the C(TABL) network with up to $10\%$ improvement (from $78.44\%$ to $88.06\%$) in average F1 measure. Compared to DAIN, the performances achieved by our normalization strategy coupled with C(TABL) or DeepLOB networks are superior to that of DAIN coupled with MLP or RNN. Regarding BN when used as an input normalization scheme, it is obvious that BN deteriorated the performance of C(TABL) networks. For example, in case of $H=10$, adding BN to C(TABL) network led to more than $10\%$ drop in averaged F1. This phenomenon is expected since BN was originally designed to reduce covariate shift between hidden layers of Convolutional Neural Network, rather than as a mechanism to normalize input time-series. 

Comparing BiN-C(TABL) with a SoTA CNN-LSTM architecture having 11 hidden layers called DeepLOB \cite{zhang2019deeplob}, it is clear that our proposed normalization layer helped a TABL network having only 2 hidden layers to significantly close the gaps when $H=10$ and $H=20$ ($81.04\%$ versus $83.40\%$ for $H=10$, and $71.22\%$ versus $72.82\%$ for $H=20$), while outperforming DeepLOB by a large margin when $H=50$ ($88.06\%$ versus $80.35\%$). 

In order to investigate how much improvement BiN can contribute to neural networks of different complexities, we evaluated BiN with a smaller TABL architecture, namely B(TABL) as proposed in \cite{tran2018temporal}. B(TABL) has only one hidden layer with a total of $5843$ parameters, compared to C(TABL) which has two hidden layers with a total of $11343$ parameters. The results are shown in Table \ref{t3}. It is clear that BiN significantly boosted both B(TABL) and C(TABL) architectures in different prediction horizons, with BiN-B(TABL) networks perform as well as BiN-C(TABL) networks in all prediction horizons, making the additional hidden layer in BiN-C(TABL) redundant. Here we should note that adding our proposed normalization layer to B(TABL) networks only leads to a mere increase of $102$ parameters while achieving the same performances as BiN-C(TABL) networks, which have approximately twice the amount of parameters. 

Since BN was proposed to normalize hidden representations, we also experimented using BiN to normalize hidden representations in TABL networks. The results are shown in Table \ref{t2}, where BiN-C(TABL) and BN-C(TABL) denote the results when BiN and BN were only applied to input, while BiN-C(TABL)-BiN and BN-C(TABL)-BN denote the results when BiN and BN were applied to both the input and hidden representations. As we can see from Table \ref{t2}, there are very small differences between the two arrangements, except a noticeable improvement for BN when the prediction horizon is $H=10$. For BiN, the this results imply that adding normalization to the hidden layers bring no additional benefit for C(TABL) networks when the input data has been properly normalized.

\begin{table}[t!]
	\begin{center}
		\caption{Comparisons between Bilinear Normalization and Batch Normalization when applied to only input layer (BiN-C(TABL) and BN-C(TABL)) or all layers (BiN-C(TABL)-BiN and BN-C(TABL)-BN}\label{t2}
		\resizebox{\linewidth}{!}{
\begin{tabular}{|c|c|c|c|c|}
\multicolumn{5}{c}{} \\ \hline
\textbf{Models}		& \textbf{Accuracy \%} 	& \textbf{Precision \%} & \textbf{Recall \%}	& \textbf{F1 \%} 		\\ \hline \hline
\multicolumn{5}{|c|}{\textit{Prediction Horizon $H=10$}} \\ \hline
BN-C(TABL)  & $79.20$	& $68.48$	&$72.36$	& $66.87$ \\ \hline
	BiN-C(TABL)	& $86.87$	& $80.29$	&$81.84$	& $\mathbf{81.04}$ \\ \hline \hline
BN-C(TABL)-BN  & $78.72$	& $68.02$	&$72.58$	& $69.98$ \\ \hline
BiN-C(TABL)-BiN & $86.84$	& $80.25$	&$81.85$	& $81.03$ \\ \hline \hline

\multicolumn{5}{|c|}{\textit{Prediction Horizon $H=20$}} \\ \hline
BN-C(TABL)  & $70.70$	& $63.10$	&$63.78$	& $63.43$ \\ \hline
	BiN-C(TABL)	& $77.28$	& $72.12$	&$70.44$	& $\mathbf{71.22}$ \\ \hline \hline
BN-C(TABL)-BN  & $71.28$	& $63.77$	&$63.65$	& $63.75$ \\ \hline
BiN-C(TABL)-BiN & $76.68$	& $71.15$	&$70.48$	& $70.80$ \\ \hline \hline
\multicolumn{5}{|c|}{\textit{Prediction Horizon $H=50$}} \\ \hline
BN-C(TABL)  	& $77.16$	& $75.70$	&$75.04$	& $75.34$ \\ \hline
	BiN-C(TABL)	& $88.54$	& $89.50$	&$86.99$	& $\mathbf{88.06}$ \\ \hline \hline
BN-C(TABL)-BN  	& $76.74$	& $75.34$	&$74.66$	& $74.97$ \\ \hline
BiN-C(TABL)-BiN	& $88.44$	& $89.36$	&$86.92$	& $87.96$ \\ \hline
\end{tabular}
	}
	\end{center}
\end{table}

\subsection{Experiments using US data}\label{us-experiments}

\subsubsection{Dataset and Experiment Setup}

While the Nordic dataset provides a reasonable testbed for our evaluation purpose, the Nordic market is less liquid compared to the US market, which is the biggest stock market worldwide. The number of intra-day orders in large-cap US stocks is significantly higher than that of the Nordic stocks, making it harder to predict the future market conditions. For the US market, we procured orders from TotalView-ITCH feed and obtained the LOB data of Amazon and Google from the 22nd of September 2015 to the 5th of October 2015. The trading hours in NASDAQ US spans from 09:30 to 16:00 (EST) and only orders submitted during this period were considered in our analysis. After the filtering process, we obtained approximately 13 millions order events for $10$ working days. Similar to the Nordic data, we used the first $7$ days for training the prediction models and the last $3$ days for testing purposes.   

In addition to forecasting the types of mid-price dynamics (up, down, stationary) at a fixed future horizon (Setting 1), we also evaluated the models in a more active setting (Setting 2), in which models were trained to predict the next movement (up or down) of the mid-price and when it occurs. That is, we have both classification (movement type) and regression (horizon value) objectives in Setting 2, with the loss function consists of the cross entropy and the mean squared error. The movement labels were derived following the same procedure used in \cite{ntakaris2018benchmark}, which includes price smoothing and movement classification based on a threshold of $0.00001$.  

For the experiments with US data, in addition to C(TABL) architecture, we also evaluated with the DeepLOB architecture \cite{zhang2019deeplob} as the predictors. Different from the Nordic dataset which was pre-normalized, the US data contains raw values for the prices and volumes. For this reason, we experimented with two static normalization methods, namely z-score normalization and min-max normalization with the results denoted as z-C(TABL) and mm-C(TABL) for C(TABL) networks, and z-DeepLOB and mm-DeepLOB for DeepLOB networks.

\begin{table}[]
\centering	
	\caption{Results for C(TABL) architecture in experiment Setting 1 of US data}
\label{t4}
\resizebox{\linewidth}{!}{%
\begin{tabular}{|l|c|c|c|c|}
\hline
\multicolumn{1}{|c|}{\textbf{Models}} & \textbf{Accuracy (\%)} & \textbf{Precision (\%)} & \textbf{Recall (\%)} & \textbf{F1 (\%)} \\ \hline
\multicolumn{5}{|c|}{\textit{Prediction Horizon $H=10$}}                        \\ \hline \hline
	C(TABL)      & $50.38$  & $41.46$  & $33.74$  & $23.62$  \\ \hline
z-C(TABL)    & $54.47$  & $50.05$  & $43.38$  & $42.50$  \\ \hline
mm-C(TABL)   & $53.13$  & $48.23$  & $40.90$  & $38.70$  \\ \hline
BN-C(TABL)   & $54.77$  & $50.20$  & $42.94$  & $41.64$  \\ \hline
DAIN-C(TABL) & $62.35$  & $60.26$  & $61.64$  & $60.62$  \\ \hline
BiN-C(TABL)  & $68.31$  & $67.03$  & $62.97$  & $\mathbf{64.31}$  \\ \hline \hline
\multicolumn{5}{|c|}{\textit{Prediction Horizon $H=20$}}                        \\ \hline \hline
C(TABL)      & $34.20$  & $37.17$  & $33.37$  & $17.74$  \\ \hline
z-C(TABL)    & $47.88$  & $47.44$  & $47.20$  & $46.45$  \\ \hline
mm-C(TABL)   & $47.37$  & $46.94$  & $46.75$  & $45.99$  \\ \hline
BN-C(TABL)   & $49.50$  & $49.29$  & $48.65$  & $47.81$  \\ \hline
DAIN-C(TABL) & $64.46$  & $64.42$  & $64.41$  & $64.40$  \\ \hline
BiN-C(TABL)  & $65.52$  & $66.15$  & $65.15$  & $\mathbf{65.26}$  \\  \hline \hline
\multicolumn{5}{|c|}{\textit{Prediction Horizon $H=50$}}                        \\ \hline \hline
C(TABL)      & $37.30$  & $36.08$  & $33.63$  & $25.83$  \\ \hline
z-C(TABL)    & $51.41$  & $50.78$  & $50.15$  & $50.23$  \\ \hline
mm-C(TABL)   & $51.71$  & $51.21$  & $49.93$  & $50.21$  \\ \hline
BN-C(TABL)   & $51.78$  & $51.37$  & $50.46$  & $50.72$  \\ \hline
DAIN-C(TABL) & $65.85$  & $63.98$  & $64.73$  & $64.25$  \\ \hline
BiN-C(TABL)  & $67.51$  & $65.98$  & $64.99$  & $\mathbf{65.38}$ \\ \hline
\end{tabular}%
}
\end{table}

\subsubsection{Experiment Results}

Table \ref{t4} shows the experiment results in Setting 1 of the US data for the C(TABL) architecture. First of all, it is clear that we obtained the worst performance when using raw data to train the predictors (results associated with C(TABL)). Between the two static normalization methods, z-score normalization exhibited better ability in preprocessing the data compared to min-max normalization. Both static normalization methods significantly improve the quality of training data. Among adaptive normalization methods, performances obtained from BN are inferior to DAIN and BiN. Overall, the proposed normalization layer when combined with C(TABL) architecture yielded the best performances in all prediction horizons compared to others. 

Table \ref{t5} shows the experiment results in Setting 1 of the US data for DeepLOB networks. Similar to the results obtained for C(TABL) networks, we also obtained the worst performance when using raw data to train the DeepLOB architecture. Between z-score normalization and min-max normalization, using the former led to slightly better results compared to the latter. While BN showed no superiority over z-score normalization, both DAIN and BiN outperformed static normalization methods. Among all normalization methods, BiN was the most suitable normalization technique to combine with the DeepLOB architecture.

\begin{table}[]
\centering	
\caption{Results for DeepLOB network architecture in experiment Setting 1 of US data}
\label{t5}
\resizebox{\linewidth}{!}{%
\begin{tabular}{|l|c|c|c|c|}
\hline
\multicolumn{1}{|c|}{\textbf{Models}} & \textbf{Accuracy (\%)} & \textbf{Precision (\%)} & \textbf{Recall (\%)} & \textbf{F1 (\%)} \\ \hline
\multicolumn{5}{|c|}{\textit{Prediction Horizon $H=10$}}                           \\ \hline \hline
DeepLOB      & $50.19$  & $31.52$  & $33.51$  & $23.28$  \\ \hline
z-DeepLOB    & $53.19$  & $44.98$  & $43.26$  & $42.21$  \\ \hline
mm-DeepLOB   & $51.83$  & $42.84$  & $39.99$  & $36.96$  \\ \hline
BN-DeepLOB   & $53.85$  & $45.78$  & $43.35$  & $42.24$  \\ \hline
DAIN-DeepLOB & $66.80$  & $64.26$  & $64.94$  & $64.54$  \\ \hline
BiN-DeepLOB  & $69.79$  & $69.82$  & $63.21$  & $\mathbf{65.05}$  \\ \hline \hline
\multicolumn{5}{|c|}{\textit{Prediction Horizon $H=20$}}             \\ \hline               \hline
DeepLOB      & $35.66$  & $23.44$  & $33.29$  & $18.47$  \\ \hline
z-DeepLOB    & $48.47$  & $47.59$  & $47.93$  & $47.36$  \\ \hline
mm-DeepLOB   & $48.46$  & $47.80$  & $47.97$  & $47.67$  \\ \hline
BN-DeepLOB   & $49.24$  & $48.14$  & $48.44$  & $47.81$  \\ \hline
DAIN-DeepLOB & $67.35$  & $67.39$  & $67.14$  & $\mathbf{67.19}$  \\ \hline
BiN-DeepLOB  & $67.50$  & $68.65$  & $66.97$  & $67.07$  \\ \hline \hline
\multicolumn{5}{|c|}{\textit{Prediction Horizon $H=50$}}      \\ \hline                      \hline
DeepLOB      & $38.62$  & $33.32$  & $33.32$  & $20.84$  \\ \hline
z-DeepLOB    & $49.85$  & $49.97$  & $49.12$  & $49.36$  \\ \hline
mm-DeepLOB   & $50.11$  & $51.57$  & $48.49$  & $49.29$  \\ \hline
BN-DeepLOB   & $50.27$  & $50.17$  & $49.73$  & $49.66$  \\ \hline
DAIN-DeepLOB & $66.86$  & $65.67$  & $65.19$  & $65.10$  \\ \hline
BiN-DeepLOB  & $67.86$  & $66.11$  & $65.56$  & $\mathbf{65.73}$ \\ \hline
\end{tabular}%
}
\end{table}

\begin{table}[]
\centering	
\caption{Results for C(TABL) and DeepLOB architectures in experiment Setting 2 of US data} 
\label{t6}
\resizebox{0.65\linewidth}{!}{%
\begin{tabular}{l|c|c|}
\cline{2-3}
             & \textbf{F1 (\%)} & \textbf{RMSE}                \\ \hline
\multicolumn{1}{|l|}{C(TABL)}         & $33.68$   & $79994377.4940$ \\ \hline
\multicolumn{1}{|l|}{z-C(TABL)}     & $53.27$   & $4118.9763$         \\ \hline
\multicolumn{1}{|l|}{mm-C(TABL)}     & $51.97$   & $110628.9429$     \\ \hline
\multicolumn{1}{|l|}{BN-C(TABL)}     & $53.57$   & $331.2658$           \\ \hline
\multicolumn{1}{|l|}{DAIN-C(TABL)}   & $51.42$   & $731.5555$           \\ \hline
	\multicolumn{1}{|l|}{BiN-C(TABL)}   & $\mathbf{54.79}$   & $\mathbf{231.4644}$             \\ \hline \hline
\multicolumn{1}{|l|}{DeepLOB}    & $41.91$   & $250.7388$            \\ \hline
\multicolumn{1}{|l|}{z-DeepLOB}   & $54.21$   & $250.7388$             \\ \hline
\multicolumn{1}{|l|}{mm-DeepLOB}  & $45.20$   & $250.7388$             \\ \hline
\multicolumn{1}{|l|}{BN-DeepLOB}  & $54.95$   & $250.7388$             \\ \hline
\multicolumn{1}{|l|}{DAIN-DeepLOB} & $32.16$   & $\mathbf{246.2643}$             \\ \hline
\multicolumn{1}{|l|}{BiN-DeepLOB} & $\mathbf{59.88}$   & $250.7388$ \\ \hline
\end{tabular}%
}
\end{table}

In experiment Setting 2, the models were trained to predict the type of the next movement of mid-price, which is measured by F1 score, as well as the horizon when it happens, which is measured by Root Mean Squared Error (RMSE). The performances of C(TABL) and DeepLOB networks using different input normalization methods are shown in Table \ref{t6}. For both network architectures, the best F1 scores were obtained using the proposed normalization method. Z-score standardization and BN performed similarly, being the second best in terms of F1 score. Min-max normalization, again, showed inferior performances compared to z-score normalization. Surprisingly, DAIN performed poorly in terms of F1 score when compared to z-score normalization in this experiment setting. Regarding the prediction of the horizon value, BiN achieved the best RMSE among all normalization methods used for the C(TABL) architecture. For the DeepLOB architecture, a peculiar phenomenon can be observed: for all normalization methods, we obtained the same RMSE, even between different runs, with DAIN as the only exception. For these models, the gradient updates toward the end of the training process seemed to only affect the classification objective and not the regression one. Even though DAIN achieved the best RMSE compared to others when applied to the DeepLOB architecture, the combination of DAIN and DeepLOB performed poorly in terms of F1 score.

From the results obtained for both Setting 1 and Setting 2, we can see that the proposed normalization method performs consistently, being the best normalization method for SoTA neural networks in most cases.

\section{Conclusions}\label{conclusions}
In this paper, we propose Bilinear Input Normalization (BiN) layer, a completely data-driven time-series normalization strategy, which is designed to take into consideration the bimodal nature of financial time-series, and aligns the multivariate time-series in both feature and temporal dimensions. The parameters of the proposed normalization method are optimized in an end-to-end manner with other parameters in a neural network. Using large scale limit order books coming from the Nordic and US markets, we evaluated the performance of BiN in comparisons with other normalization techniques to tackle different forecasting problems related to the future mid-price dynamics. The experimental results showed that BiN performed consistently when combined with different state-of-the-arts neural networks, being the most suitable normalization method in the majority of scenarios. 

\section{Acknowledgement}
The authors wish to acknowledge CSC – IT Center for Science, Finland, for computational resources.

\bibliography{reference}
\bibliographystyle{ieeetr}

\end{document}